\documentclass[twocolumn,preprintnumbers,amsmath,amssymb,nofootinbib,superscriptaddress]{revtex4-1}

\usepackage{epigraph}
\usepackage{array}
\usepackage{braket}
\usepackage{xfrac}
\usepackage{bbold}
\usepackage{bbm}
\usepackage{array}
\usepackage{enumerate}
\usepackage{amsmath}

\usepackage{hyperref}

\usepackage{graphicx}
\usepackage[font=small]{caption}
\usepackage{subcaption}

\usepackage{amsfonts}
\usepackage{amssymb}
\usepackage{gensymb}

\usepackage[font={small,it}]{caption}

\renewcommand\citet[1]{Ref.~\cite{#1}}

\newcommand\effect[1]{\reflectbox{\ensuremath{\vec{\reflectbox{\ensuremath{#1}}}}}}

\newcommand{\half}[1][1]{\frac{#1}{2}}
\newcommand{\recip}[1]{\frac{1}{#1}}
\newcommand{\ketbra}[2]{\ket{#1}\!\bra{#1}}

\newcommand{\trace}[1]{\text{Tr}\left(#1\right)}
\newcommand{\commutator}[2]{\left[#1,\;#2\right]}
\newcommand{\pderivative}[2]{\frac{\partial #1}{\partial #2}}
\newcommand{\expect}[1]{\langle #1\rangle}

\newcommand{\ident}{\mathbb{1}}

\newcommand{\SU}[1]{\mathrm{SU}\mkern-1mu(#1)}
\newcommand{\SO}[1]{\mathrm{SO}\mkern-1mu(#1)}

\newcommand{\inlineheading}[1]{\noindent\textbf{#1---}}

\begin{document}

\title{On defining the Hamiltonian beyond quantum theory}

\author{Dominic Branford}
\affiliation{Department of Physics, University of Warwick,
Coventry, CV4 7AL, United Kingdom}
\affiliation{Atomic and Laser Physics, Clarendon Laboratory,
University of Oxford, Parks Road, Oxford, OX1 3PU, United Kingdom}

\author{Oscar C. O. Dahlsten}
\email{dahlsten@sustc.edu.cn}%
\affiliation{Southern University of Science and Technology (SUSTech), Nanshan District, Shenzhen, China}
\affiliation{Atomic and Laser Physics, Clarendon Laboratory,
University of Oxford, Parks Road, Oxford, OX1 3PU, United Kingdom}
\affiliation{London Institute for Mathematical Sciences,
35a South Street, Mayfair, London, W1K 2XF, United Kingdom}

\author{Andrew J. P. Garner}
\affiliation{Institute for Quantum Optics and Quantum Information,\\
Austrian Academy of Sciences, Boltzmanngasse 3, A-1090 Vienna, Austria}
\affiliation{Centre for Quantum Technologies, National University of Singapore, 3 Science Drive 2, 117543, Singapore}
\affiliation{Atomic and Laser Physics, Clarendon Laboratory,
 University of Oxford, Parks Road, Oxford, OX1 3PU, United Kingdom}

\date{August 14, 2018}

\begin{abstract}
Energy is a crucial concept within classical and quantum physics.
An essential tool to quantify energy is the Hamiltonian.
Here, we consider how to define a Hamiltonian in general probabilistic theories---a framework in which quantum theory is a special case. 
We list desiderata which the definition should meet. 
For 3-dimensional systems, we provide a fully-defined recipe which satisfies these desiderata. 
We discuss the higher dimensional case where some freedom of choice is left remaining. 
We apply the definition to example toy theories, and discuss how the quantum notion of time evolution as a phase between energy eigenstates generalises to other theories.
\end{abstract}

\maketitle

\section{Introduction}

\epigraph{The central conception of all modern physics is the ``Hamiltonian"
}{\textit{Erwin Schr\"odinger~\cite{Hankins1980,Coopersmith2010}}}

Energy is a central concept in classical and quantum physics, and hence plays a significant part in all of physical science~\cite{Hankins1980,Coopersmith2010}.
The dynamical behaviour of mechanical systems can be understood in terms of energy conservation~\cite{arnold_mathematical_1989}.
Without energy, there can be no thermodynamics.
Meanwhile, quantizing the allowed energies of electrons trapped around nuclei gives rise to atomic structure, leading to the periodic table and chemistry~\cite{cohen-tannoudji_quantum_1977}.
For a physical theory to be practically useful, it seems it should possess an analytic tool for the systematic treatment of energy.

The {\em Hamiltonian} is one such tool in classical and quantum physics, serving a dual role: 
(i) It is a physical observable---some measurable quantity---that acts as a conserved quantity, 
 determining conserved quantities in classical mechanics and good quantum numbers in quantum mechanics through the Poisson bracket and commutator respectively.
(ii) It is the generator of time evolution---it dictates the dynamical behaviour of systems changing with time.
These two roles particularly come together in thermodynamics, wherein it functions as a conserved quantity in the first law,
 and also defines the natural thermal state that systems tend towards (in the form of the Boltzmann distribution, whose form is similar to that of Hamiltonian evolution, but in \mbox{imaginary} time).

There is a recent paradigm that views quantum theory as a special case in a wider framework of theories known as the {\em convex framework} 
or {\em generalised probabilistic theories} (GPTs)~\cite{hardy_quantum_2001,barrett_information_2007,masanes_derivation_2011,janotta_generalized_2014}.
Within this framework one can formalise alternative theories that go beyond quantum theory (e.g.\ by allowing for states with stronger correlations than the maximally entangled quantum states~\cite{popescu_quantum_1994}).
This framework attracts significant foundational interest for a variety of motivations, including:
(i) to derive quantum theory from natural axioms by firstly considering a wider set of theories and then applying axioms to rule all but one out~\cite{hardy_quantum_2001,hardy_reformulating_2011, masanes_derivation_2011, dakic_quantum_2011,chiribella_informational_2011,barnum_higher-order_2014},
(ii) to identify the specific features of quantum theory responsible for particular phenomena~\cite{barnum_cloning_2006,chiribella_entanglement_2016}
(iii) to understand the foundational basis of quantum cryptography:  
 by its operational and data-focussed nature, the convex framework allows for the stripping away of unnecessary concepts. 
This makes it easier (for example) to consider device-independent cryptography~\cite{barrett_no_2005,acin_bells_2006,acin_device-independent_2007}.

At its core, the convex framework is information-theoretical, and data-driven: it focuses primarily on observable statistics, rather than the underlying mechanisms generating them.
There is thus interest in developing the connections between this framework and traditional physical principles.
Among many examples, consideration has been given as to what is required to arrive at thermodynamical concepts within this framework~\cite{short_entropy_2010,barnum_entropy_2010,kimura_distinguishability_2010,muller_unifying_2012, chiribella_entanglement_2015,barnum_entropy_2015,chiribella_operational_2015,kimura_entropies_2016, krumm_thermodynamics_2017,chiribella_microcanonical_2017}, 
what sort of particles could exist in alternative theories~\cite{dahlsten_particle_2013},
what interference fringes might we see~\cite{sorkin_quantum_1994, ududec_three_2011, garner_framework_2013, dahlsten_uncertainty_2014, barnum_higher-order_2014, garner_quantum_2017, lee_higher-order_2017, barnum_ruling_2017}, which theories allow for tunnelling through energy barriers~\cite{lin_tunnelling_2016}, and the potential computational power of machines in certain GPTs~\cite{lee_computation_2015,lee_generalised_2016,lee_bounds_2016,lee_deriving_2016,barrett_computational_2017,lee_oracles_2017,garner_2017}.

Exciting recent results from \citet{barnum_higher-order_2014} have shown the value of introducing the concept of the Hamiltonian as both an observable and generator of evolution in order to select quantum mechanics among other rival theories.
Starting from other axioms (that limit the theories to Jordan algebras), they use the existence of such an energy observable assignment to single out quantum theory.
Meanwhile, in \citet{chiribella_entanglement_2016}, information-theoretic axioms are used to recover key properties of statistical mechanics, such as entropy and Landauer's principle.

However, there is still an important piece missing in this field that we aim to address here: 
 namely how to concretely define Hamiltonians in a wide range of convex theories other than quantum theory.
Here, we do not attempt to rederive quantum theory (and so do not seek a complete list of axioms as per \citet{barnum_higher-order_2014} that would single out quantum theory), 
 but instead list a set of desiderata that we believe any reasonable definition of a Hamiltonian should have.
By doing so, we provide a firmer base to enable progress in a wide range of questions involving energy.

We proceed as follows:
 we review the role of the Hamiltonian in quantum theory, 
 taking a geometric approach in which the Hamiltonian can be represented as a real vector and the axis of rotation. 
We list and justify desiderata for a Hamiltonian definition in general theories, and analyse their implications. 
Focusing on 3-dimensional theories\footnote{
By $n$-dimensional theories, we mean those where normalized states are determined by $n$ real degrees of freedom.
}, we give an explicit recipe which satisfies the desiderata, and apply this recipe to examples. 
Finally, we discuss the higher dimensional case, 
and how the time evolution relates to {\em phase transformations} associated with energy eigenstates.  

\section{Geometric representation of Hamiltonian dynamics}
\label{sec:quantum_real}
In quantum mechanics, the reversible time evolution of a state $\rho$ of a system is determined by the Hamiltonian $H$ according to the von Neumann equation
\begin{equation}\label{vonneumanneq}
	\frac{d\rho}{dt} = - \frac{i}{\hbar} \left[H,\rho\right],
\end{equation}
which for a time-independent $H$, has the well-known \mbox{solution} 
\begin{equation}
	\rho(t) = \exp\left({-i\frac{Ht}{\hbar}}\right) \rho(0) \exp\left({i\frac{Ht}{\hbar}}\right),
\end{equation}
for initial state $\rho\!\left(0\right)$ evolving over time $t$.

\subsection{Two-level quantum system}
In the simplest case---a two-level quantum system (qubit)---we can interpret this evolution geometrically using the Bloch sphere.
A general normalized qubit state $\rho$ can be expressed in the form
\begin{equation}\label{statequbitvec}
\rho=\half\mathbb{1}+\half\sum\limits_{i=1}^3 u_i\sigma_i,
\end{equation}
where $\{u_i\}$ are real parameters satisfying $\sum_i |u_i|^2 \leq 1$ (the uncertainty principle\footnote{
Specifically, from the Robertson-Schr\"odinger uncertain relation~\cite{robertson_1929,schrodinger_1930} as applied to the three Pauli operators -- see, for instance, appendix A of~\cite{garner_2017}
}),
 and $\sigma_i$ are the Pauli matrices.
The corresponding definition for a Hamiltonian is
\begin{equation}\label{Hamiltonianqubitvec}
H=\half v_0\mathbb{1}+\half\sum\limits_{i=1}^3 v_i\sigma_i,
\end{equation}
where $\{v_i\}$ are real numbers. 

The Pauli matrices, which---along with the identity---form a basis for the $2 \times 2$ Hermitian matrices, satisfy 
\begin{gather}
\label{paulitrace}
\trace{\sigma_i\sigma_j}=2\delta_{ij},\\
\label{paulicommutation}
\commutator{\sigma_j}{\sigma_k} = 2i \epsilon_{jkl} \sigma_l,
\end{gather}
where $\epsilon_{jkl}$ is the (totally anti-symmetric) Levi-Civita symbol, 
 which corresponds to the {\em structure constants} of the $\SU{2}$ algebra~\cite{hioe_n-level_1981}.

\begin{figure}[h]
\centering
\includegraphics[width=0.35\textwidth]{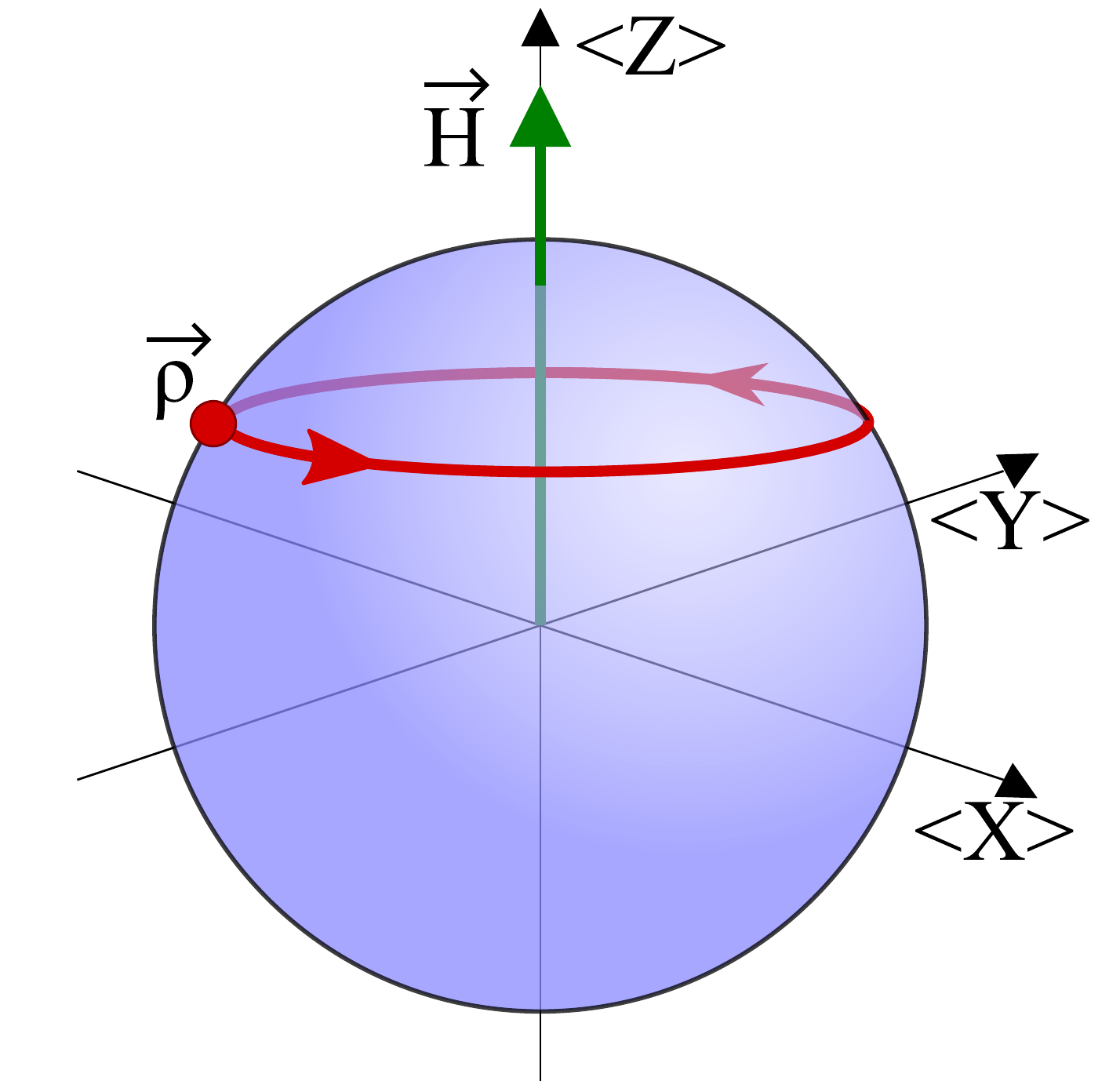}
\caption{{\bf The Hamiltonian in quantum theory} (green arrow) acts as the axis of rotation as states (such as that represented by the red dot) evolve in the Bloch sphere.
}
\label{fig:rotate}
\end{figure}

As structure constants are important in what follows, we recall their definition:  
 given a set of basis vectors $\vec{e_i}$ for the underlying vector space of the algebra, 
 the structure constants (or coefficients) $f_{ijk}$ express the multiplication $\circ$ of pairs of vectors as a linear combination:
\begin{equation}
	\vec{e}_{i} {\;\circ\;} \vec{e} _{j}=\sum _{k} f_{ijk}\vec{e}_{k}.
\end{equation}
The above allows us to define real vectors $\vec{\rho}$ (the Bloch vector) and $\vec{H}$ that represent the state and Hamiltonian respectively as $H_i=\trace{H\sigma_i}$ and $\rho_i=\trace{\rho\sigma_i}$ with $i\in\{0,1,2,3\}$.
Here, $\rho_0$ is the normalisation~\cite{dahlsten_uncertainty_2014}, and $H_0$ is the average of energy over all states, and neither plays any dynamical role.
Combining the above equations allow us to express the qubit evolution as
\begin{equation}
\label{eq:crossprod}
\pderivative{\vec{\rho}}{t}=-\recip{\hbar}\vec{\rho}\times\vec{H},
\end{equation}
 where $\times$ is the vector cross product (note: for $i,j,k\in\{1,2,3\}$, $f_{ijk}$ and the Levi-Civita symbol $\epsilon_{ijk}$, by which the cross product is usually denoted, are one and the same).  
Thus, in the Bloch sphere representation we can describe the state evolution as a rotation around the Hamiltonian axis (particularly, the axis between the eigenstates of the Hamiltonian, often referred to as the {\em energy eigenstates}). 
This rotation has a speed dependent on the magnitude of the Hamiltonian $\vec{H}$.
This is depicted in Fig.~\ref{fig:rotate}. 

\subsection{General finite-dimensional case}
The state and Hamiltonian vector definitions generalise to a $d$-level system by way of the generalised Bloch vector~\cite{hioe_n-level_1981,kimura_bloch_2003,bengtsson_geometry_2006,bertlmann_bloch_2008}.
As well as normalization constraints, there are other restrictions on which generalised Bloch vectors correspond to valid quantum states (i.e.\ to prevent the corresponding density matrix having negative eigenvalues~\cite{jakobczyk_geometry_2001}).
For a general quantum system we write:
\begin{align}
\label{eq:QuantumRealRep}
\rho=\recip{d}\mathbb{1}+\half\sum\limits_{j=1}^{d^2-1}u_j\lambda_j,\\
H=\half[v_0]\mathbb{1}+\half\sum\limits_{k=1}^{d^2-1}v_k\lambda_k,
\end{align}
where $\{\lambda_i\}$ are the generalised Gell-Mann matrices~\cite{bertlmann_bloch_2008}, which form a basis of Hermitian operators much like the Pauli matrices.
The generalised Gell-Mann matrices also satisfy generalizations of \eqref{paulitrace} and \eqref{paulicommutation}, such that $\trace{\lambda_i \lambda_j} = 2\delta_{ij}$ and $[\lambda_j,\lambda_k] = 2i\,f_{jkl}\,\lambda_l$ where $f_{jkl}$ are the structure constants of the $\SU{d}$ algebra~\cite{hioe_n-level_1981}.
One can then identify $(d^2\!-\!1)$--dimensional real-vectors $\vec{\rho}$ and $\vec{H}$ 
 with elements $\rho_i=\trace{\rho\lambda_i}$ and $H_i=\trace{H\lambda_i}$ for $i\in\{0,1,2,\cdots,d^2-1\}$,
 respectively describing the state and Hamiltonian of a $d$-dimensional quantum system.

The state evolution follows the von Neumann equation (eq.~\eqref{vonneumanneq}) written in matrix form:
\begin{align}\nonumber
\pderivative{\rho}{t}&= - \frac{i}{\hbar} \commutator{\half[v_0]\mathbb{1}+\half\sum\limits_{k=1}^{d^2-1}v_k\lambda_k}{\recip{d}\mathbb{1}+\half\sum\limits_{j=1}^{d^2-1}u_j\lambda_j}\\
&=\recip{2\hbar}f_{ijk}u_iv_j\lambda_k,
\label{eq:qudit_evolution}
\end{align}
where we use implicit summation notation.
From this we can extract the vector form of this equation by taking the trace of Eq.~\eqref{eq:qudit_evolution} multiplied by \( \lambda_l \) to obtain 
\begin{equation}
	\dot{u}_l = \frac{1}{\hbar} f_{ijl} u_i v_j.
\end{equation}

\section{Beyond quantum theory: \mbox{The Convex Framework}}
We shall employ aspects of the convex framework to go beyond quantum theory~\cite{hardy_quantum_2001,barrett_information_2007,masanes_derivation_2011,janotta_generalized_2014}.
In this framework, a theory is defined by 
 (i) a finite-dimensional, compact convex set of normalized states, 
 (ii) a convex set of effects (measurement-outcome pairs), 
 and (iii) a set of transformations.
States may be represented by real vectors. 
Effects are linear functionals that map states to probabilities associated with a particular outcome given a measurement,
 and these can also be represented by real vectors.
Particularly, the state $\vec{\rho}$ and the measurement-outcome effect $\effect{e_i}$ can be represented such that the probability $p_i$ of a measurement yielding a given outcome $i$ on a particular state is calculated as the standard Euclidean inner product ($\cdot$) between the two vectors:
\begin{equation}
p_i=\effect{e_i}\cdot \vec{\rho}.
\end{equation}

The set of normalized states generate a positive convex cone {$\Omega$} of (in general) unnormalized states. 
The apex of this cone represents an ``absent'' state on which every effect returns $0$ probability.
If $\vec{\rho_1}$ and $\vec{\rho_2}$ are in the set of allowed states, then so is the state $\rho = p \vec{\rho_1} + (1-p)\vec{\rho_2}$ for $p\in[0,1]$. 
This convex combination is interpreted as a probabilistic preparation of a state, just as in quantum theory. 
Likewise, the set of effects also form a positive cone, but in general this is a different cone from that of states (only coinciding in special cases; notably, quantum theory where every bra has a unique associated ket).

There is a special unique effect known as the {\em unit effect}, $\effect{u}$, 
 such that $\effect{u} \cdot \vec{s} = 1$ for every normalized state $\vec{s}$ in the theory\footnote{
 Geometrically, the set of normalized states correspond to a hyperplane intersection with the cone of all states. 
}.
The unit effect also allows the definition of a {\em measurement} within a theory: a collection of effects $\{\effect{f_i}\}_{i=1\ldots N}$ such that summing $\sum_{i=1}^N \effect{f_i} \cdot \omega = u\cdot \omega $ for all states $\omega \in \Omega$.
This corresponds to the assigned probabilities of all effects within this set summing to unity for any choice of normalized state.
(Assuming that all such measurements can actually be made on a system amounts to a strong version  of the {\em no-restriction hypothesis}~\cite{chiribella_2010,janotta_2013}).

As the Hamiltonian is an observable rather than a measurement outcome, we shall also use the notion of a vector representing an observable. 
Consider a measurement $X$ on a state $\rho$ (with real vector representation $\vec{\rho}$), 
 where each measurement outcome $x_i$ occurs with probability $p_i\!\left(\rho\right) = \effect{f_i}\cdot\vec{\rho}$.
In general, we want the expectation value of $X$ with respect to any given state $\rho$ to satisfy
\begin{equation}
\langle X \rangle_{\rho} = \sum_i p_i\!\left(\rho\right) x_i.
\end{equation}
This motivates the definition of an observable $\effect{X}$ as a linear combination of effects\footnote{
This definition follows similar definitions in~\cite{garner_phase_2015, chiribella_entanglement_2016}.
In general, an observable can be treated as a functional that maps states onto a number, for example representing an expectation value of a measurement, as per \cite{barnum_higher-order_2014}.
}
\begin{equation}
\effect{X}:=\sum_i x_i \effect{f_i},
\label{eq:observable}
\end{equation}
(independent of the state $\rho$)
such that 
\begin{equation}
\langle X \rangle_{\rho} = \effect{X}\cdot\vec{\rho}
\end{equation}
for every $\rho$.

The evolution of the state vector is a linear map (i.e.\ can be represented as a matrix) that takes states to states,
 following from the postulate that the map should not depend on the probabilities of the states involved~\cite{hardy_quantum_2001,barrett_information_2007}. 
Thus, if the state of a system evolves in time $t$, it can be written as 
\begin{equation}
\label{eq:timematrix}
\vec{\rho}(t)=M(t)\vec{\rho}(0),
\end{equation}
where $M(t)$ is some real matrix and $\vec{\rho}(0)$ is the initial state.

Transformations can be composed such that application of $M_1$ followed by $M_2$ corresponds to the matrix product $M= M_2 M_1$,
 and there exists the possibility of doing nothing to the state at all (corresponding to identity matrix $\ident$).
As such, the set of all transformations in a theory forms a monoid.
Particularly, the set of transformations that can be reversed (that is where for every $M$ in the set, there exists an $M'$ in the set such that $M' M = \ident$),
 forms a group $\mathcal{T}$ which will be a subgroup of the automorphism group of the theory's state space (or the automorphism group itself).

Broadly, one can consider the {\em reversible dynamics} of a theory as a single parameter subgroup of $\mathcal{T}$, where the single parameter $t$ corresponds to the passage of time.
In this article, we shall restrict our discussion to time-independent dynamics (such as those induced by the time-independent Schr\"{o}dinger equation in quantum theory).
Here, one may consider {\em continuous dynamics} as a homomorphism%
\footnote{
Without time-independence, this map is not necessarily homomorphic. 
E.g.\ composition of evolutions for $3$ seconds and for $5$ seconds is equivalent to evolving the system for $8$ seconds only if ``$5$ seconds of evolution'' corresponds to the same group element whether the evolution begins at time $t=0$ or at time $t=3\;{\rm seconds}$.
} 
 from the real numbers under addition (viz.\ the passage of time) to an element of the transformation group: $\mathbb{R} \to \mathcal{T}$.
{\em Discrete dynamics} (discussed further in Section~\ref{sec:discrete}) correspond to quantized time, and so act on a more restricted domain to map $\mathbb{Z} \to \mathcal{T}$.
The specification of these maps (if they exist) is a property of the particular theory.

Classical probability theory fits into the above formalism, wherein the states are probability vectors, the effects are the natural basis of those vectors, and normalization-preserving transformations are stochastic matrices.
Quantum theory also fits into this formalism, 
 much as described in Sec.~\ref{sec:quantum_real}:
 states can be expressed as a Bloch vector, observables as another vector (e.g.\ like $\vec{H}$), and the transformations as linear maps taking valid  states to valid states.

\section{Desiderata for Hamiltonians beyond quantum theory}
\label{sec:desiderata}
To what extent can one define a Hamiltonian in probabilistic theories other than quantum theory? 
We shall make reference to the following possible desiderata of the definition of the Hamiltonian:
\begin{enumerate}[A.]
\item The Hamiltonian should be an observable [{\bf OBS}].
\item The Hamiltonian should (at least partially) determine the generator of time evolution [{\bf GEN}]. 
\item The expected value of the Hamiltonian should be invariant under time evolution [{\bf INV}].
\item The Hamiltonian should be consistent with the quantum definition [{\bf QUAN}]: Given quantum theory in the form of states and reversible transformations, the definition should give the same pairs of Hamiltonians and time evolutions as standard quantum theory would.
\end{enumerate}

In~\citet{barnum_higher-order_2014} (particularly, Definition 30), the existence of an {\em energy observable assignment} 
 is taken to mean that there is a mapping between energy observables ({\bf OBS}) and generators ({\bf GEN})
 that is one-to-one, respects property {\bf INV},
 and assigns different values to at least 2 different states whenever the time-evolution is non-trivial.
Thus, their definition of an energy observable assignment is consistent with our desiderata for a Hamiltonian.

In particular, we shall discuss example theories (such as box world~\cite{barrett_information_2007}, and Spekkens' toy model~\cite{spekkens_evidence_2007}) that would have already been ruled out by the other axioms assumed in~\citet{barnum_higher-order_2014} (indeed, as is a key point of~\cite{barnum_higher-order_2014}: from a restricted set of theories that respect their earlier axioms, the {\em only} theory respecting their energy observable assignment is quantum theory).

\subsection{Implications of {\bf OBS}}
\label{sec:OBS}
In line with Eq.~\ref{eq:observable}, the Hamiltonian can be represented as the observable
\begin{equation}
\label{eq:EnergyObservable}
\effect{H}=\sum_i H_i\effect{h_i},
\end{equation}
where $\effect{h_i}$ are effects forming a measurement such that $\sum_i \effect {h_i}\cdot \vec{\rho}=1 $ for all normalised states $\vec{\rho}$.

In general, this is not a unique decomposition: a given $\effect{H}$ could have many different sets of $\{e_i\}$ that satisfy Eq.~\eqref{eq:EnergyObservable}. 
The standard way of decomposing an observable in quantum theory---the spectral decomposition---is associated with orthogonal projectors $\ketbra{i}{i}$.
An operational significance of orthogonal projectors is that they are states that can be deterministically distinguished in a single measurement.
For example, the Pauli matrix $Z$ is usually written as $Z=\ketbra{0}{0}-\ketbra{1}{1}$ in terms of effects $\ketbra{0}{0}$ and $\ketbra{1}{1}$ associated with eigenvalues $+1$ and $-1$ respectively.

However, in general theories the existence of a unique spectral decomposition is not guaranteed. 
Rather this is a property that can be taken as an axiom towards selecting quantum theory~\cite{chiribella_entanglement_2016, wilce_2016, krumm_thermodynamics_2017}.
As such, an alternative statement {\bf OBS*} that the Hamiltonian can be represented as a {\em spectral observable} would be stronger than {\bf OBS}, but would preclude the finding of Hamiltonians in many non-quantum theories.

\subsection{Implications of {\bf GEN}}
In quantum mechanics, state spaces have continuous symmetries,
 while many other state spaces of interest in the GPT framework feature discrete symmetries~\cite{barrett_information_2007,spekkens_evidence_2007}. 
When discussing the time evolution, this introduces a potential issue: 
 the continuous symmetries of the quantum state spaces are needed to allow unitary evolution over an arbitrary time step.
For instance,
 consider a cubic state space ($3$-in $2$-out gbit~\cite{barrett_information_2007}) evolving 
 under quantum dynamics with Hamiltonian $Z$. 
States where $\langle X \rangle^2 + \langle Y \rangle^2 > 1$ (strictly) are not mapped to gbit states for all times, 
 but only for times that induce rotations of $n\pi/2$ where $n\in\mathbb{Z}$.
Since restriction to discrete rotations has somewhat qualitatively different physical implications, we shall discuss the continuous case here, and defer discussion of the discrete case until Section~\ref{sec:discrete}.

For a generic transformation to be a {\em time evolution},
 the concept of time must be involved: namely that the transformation takes the state of the system at one time to a state at another.
Moreover, one might wish to admit the concept of performing the ``same type'' of transformation, over different lengths of time.
For this structure to be present, we would then need to identify one-parameter subgroups of transformations, wherein this single parameter corresponds to time.
{\bf GEN} then states that it is the Hamiltonian within a theory that should determine the particular one-parameter subgroup by which a system evolves.

Normalized state-spaces are topologically bounded.
The reversible linear transformations on them hence form a compact Lie group (or some finite subgroup thereof)~\cite{masanes_derivation_2011}.
Thus there exists a unitary---and moreover, for real vector spaces, orthogonal---representation of these transformations~\cite{Simon96}.
Orthogonal matrices have determinant $\pm 1$, and eigenvalues with absolute value $1$.
For theories with continuous time, we must restrict ourselves to elements of $O(n)$ that are connected to the identity.
This selects {the {\em special orthogonal} matrices $\SO{n}$, with determinant $+1$.
We can see this by considering the orthogonal matrix $\mathcal{O}(\tau)$ that evolves the system over time $\tau$,
 and requiring from continuity, the existence of some evolution for half that time, $\mathcal{O}(\tau/2)$, defined such that when it is applied twice $\mathcal{O}(\tau/2) \mathcal{O}(\tau/2) = \mathcal{O} (\tau)$.
If $\mathrm{det}[\mathcal{O}(\tau)]=-1$, 
 then $\mathcal{O}(\tau/2)$ cannot have determinant $\pm 1$, since $\det(PQ)=\det(P)\det(Q)$, and hence cannot be a valid transformation.

When the transformation is given by a special orthogonal matrix,
 we may consider the infinitesimal change in state $\vec{\rho}$ between times $t$ and $t+{\rm d}t$:
\begin{equation}
\label{eq:Infinitesimal}
\vec{\rho}\left(t+ {\rm d}t\right) = \left[\ident +  {\rm d}t\, A\!\left(t\right)\right] \vec{\rho}\left(t\right),
\end{equation}
where $A$ is known as the {\em generating matrix}.
For time-independent%
\footnote{
For time-dependent $A$, one integrates Eq.~\eqref{eq:Infinitesimal} into a time-ordered exponential:
$M(\tau) = \mathcal{T}\left\{\exp\left[\int_0^\tau A(t)\, {\rm d}t \right]\right\}$, 
 where $\mathcal{T}(\cdot)$ denotes that every term in the expansion of the exponent only appears in increasing time order. 
This explicit ordering is necessary since in general $A(t)$ and $A(t')$ might not commute at different times.
}
 $A$, this relates to the finite transformation $M(t)$ by way of the exponential map 
\begin{equation}
\label{eq:ExpMap}
M(t) = \exp\left(A t \right).
\end{equation}

As a generator of the special orthogonal group ${\rm SO}(d)$ (where $d$ is the dimension of the set of normalised states) $A$ is anti-symmetric: $A_{ij}=-A_{ji}$
 and hence has $d(d-1)/2$ degrees of freedom in general.

With this in mind, let us thus consider the implications of {\bf GEN}.
Namely, if the Hamiltonian $\effect{H}$ should (partially) determine the generator of time evolution,
 it would be sufficient to describe this $A$ as being dependent on $\effect{H}$.
In keeping closer to quantum mechanics one could place the stronger requirement that $A$ be a linear function of $\effect{H}$.
The exact form of this dependency will be constrained by other desiderata -- in particular, {\bf INV}.

Finally, let us consider whether or not the Hamiltonian must {\em uniquely} determine the generator of time evolution.
This would be a stronger requirement (say {\bf GEN*}), and would accordingly limit the number of theories that could admit Hamiltonians.
Indeed, just from na\"ive dimension counting, matching the $d$ parameters in the observable with the $d(d-1)/2$ degrees of freedom in the generator posits a challenge (the inability to do so is used in \cite{barnum_higher-order_2014} to rule out higher-dimensional alternatives to the Bloch sphere\footnote{
\Citet{barnum_higher-order_2014} use an inverse statement of {\bf GEN}, and consider the ``observability of energy'' as a postulate for quantum theory. 
Namely, they specify that the generator of dynamics should be able to uniquely determine an observable. 
On top of a set of axioms that restricts theories to Jordan algebras, this uniquely singles out quantum theory.
}).

In the case of finite-dimensional quantum mechanics this constraint comes through the dynamics being governed by the $\SU{d}$ group as in Eq.~\eqref{eq:qudit_evolution}.
In general theories, there is no guarantee of a one-to-one relation linking time evolution with the energy observable.
Indeed, it could be that the Hamiltonian can only identify the dynamics up to a subgroup with more than a single parameter of freedom (i.e.\ more than just time).
Here, perhaps another observable would make up the deficit, and completely determine the system's dynamics;
 or perhaps otherwise we may have a priori restrictions on the theory's dynamics such that given specification of such a sub-group by the Hamiltonian, a particular one-parameter subgroup is systematically chosen.
One might find it physically motivated to avoid such situations and hence require {\bf GEN*}. 
This would limit one's ability to find Hamiltonians, except in theories similar to quantum theory.

\subsection{Implications of INV}
\label{subsec:INV}
If we impose {\bf INV}, such that the energy expectation value is conserved, we require:
\begin{equation}
\effect{H}\cdot\vec{\rho}(t)=\effect{H}\cdot e^{A t'}\vec{\rho}(t),\, \forall t',\rho.
\end{equation}
We can perform a Taylor expansion for $e^{A t'}$, which leads us to the equality
\begin{equation}
H_iA_{ij}(t'\mathbb{1}_{jk}+\half t'^2A_{jk}+\recip{6}t'^3 A_{jl} A_{lk}\cdots)\rho_k=0,\,\forall t',\rho.
\end{equation}
In order to satisfy this for an arbitrary state at any time $t'$, we seek a solution to $H_i A_{ij}=0$, which places $d$ constraints upon $A$.
If $A$ is non-zero (i.e.\ for non-trivial dynamics), then to solve this for arbitrary states $A$ must have some dependence on $\effect{H}$.
If this dependence is linear then we can write $A_{ij} = g_{ijk} H_k$, where for non-trivial dynamics $H_k \neq 0$ for at least one $k$. 
(The case of $H_k = 0\;\forall\;k$ trivially satisfies {\bf INV}).
Then, anti-symmetry $g_{ijk}=-g_{kji}$ is sufficient to satisfy $H_i A_{ij}=0$ since $H_i g_{ijk} H_k = H_k g_{kji} H_i$.

In principle $g_{ijk}H_i H_k=0$ could be obtained through other solutions, but we seek a general mechanism that satisfies this for independent choice of $\effect{H}$.
The tensor $g_{ijk}$ can be decomposed into a part symmetric under $i \leftrightarrow k$, and a part anti-symmetric under the same as $g_{ijk}=g^{\mathrm{S}}_{ijk} + g^{\mathrm{A}}_{ijk}$.
Suppose $g_{ijk}H_i H_k = 0$ for some $\effect{H}$.
For the Hamiltonian $\effect{H}'$ where $H'_m = H_m + c_m$, we would require $g_{ijk}H'_i H'_k=0$ also.
Subtracting $g_{ijk}H_i H_k = 0$  from this, we require $g_{mjm}c_m^2 + g_{ijm} c_m H_i + g_{mjk} c_m H_k = 0$.
Any $g^{\mathrm{A}}_{ijk}$ satisfies this, while the symmetric part requires $g^{\mathrm{S}}_{mjm} c_m^2 + 2 g^{\mathrm{S}}_{ijm} c_m H_i = 0$, which cannot be satisfied without introducing a dependence of $g_{ijk}$ on $\effect{H}$.
Thus, for $g_{ijk}$ to generally satisfy {\bf INV} it must {\em necessarily} be anti-symmetric under $i \leftrightarrow k$.
The addition of the previous restriction $A_{ij}=-A_{ji}$ then renders $g_{ijk}$ a totally anti-symmetric tensor. 
 This result leads to the same expression as earlier: 
\begin{equation}\label{eq:generalev}
\dot{\rho}_i=g_{ijk}\rho_j H_k.
\end{equation}
Looser constraints may be possible by significantly constraining the allowed Hamiltonians and states.

In general $A$ need not even be linear in $\effect{H}$; a more general form $A_{ij}=g_{ijkm}{(H_k)}^m$ could be constructed.
However, in this more general picture the same argument imposing $g_{ijkm}=-g_{kjim}$ would be insufficient to conserve the expectation of $\effect{H}$.

Note that conserving an operator's quantity is not in general sufficient to single out $H$.
Even in the quantum case 
 the Heisenberg equation $i \hbar \frac{d\hat{O}}{dt} = - [ \hat{H}, \hat{O} ]$ implies that {\em any} observable $\hat{O}$ that commutes with Hamiltonian $\hat{H}$ is conserved. ($\hat{H}$ itself is trivially conserved by this equation).
This leads naturally to the definition of ``good quantum numbers'' such as angular momentum in atoms or the spin of particles.

\subsection{Implications of QUAN}
{\bf QUAN} means that in the case of quantum states and quantum generators, a generic recipe for producing Hamiltonians should give the same equation for the Hamiltonian as quantum theory would give. 
In particular Eq.~\ref{eq:crossprod} and the higher dimensional equivalent Eq.~\ref{eq:qudit_evolution} should follow from the definition in this case.  

\section{Discrete evolution}
\label{sec:discrete}
In order to widen the reach of theories that might meaningfully have a Hamiltonian, we now now consider the possibility of defining Hamiltonians for systems where the evolution is governed by discrete dynamics.

First, we remark that desideratum {\bf OBS} is unaffected by the dynamics.

Next, we consider to what extent does desideratum {\bf GEN} makes sense in the discrete context.
Discrete transformations (particularly rotations and reflections) are finite subgroups of the (infinite) Lie groups that govern continuous transformation. 
For a continuous transformation, infinitesimal generators can be integrated to an arbitrary time to produce the group element that corresponds to the desired action on the state space.
For discrete rotations the same infinitesimal generators may also be used, but to ensure that a valid group element produced one must integrate only to allowed times from a discrete set.
Symmetry dictates that these times will have the form $n\tau$ where $n\in\mathbb{Z}$ and $\tau$ is a fixed time determined by the form of the Hamiltonian. 
(Setting the value of $\tau$ will be discussed further in Section~\ref{sec:energyeigenstates}.)

Additionally, if we permit non-continuous evolution -- and especially do not require that every evolution can be divided into an evolution over a shorter time period -- then the theory could theoretically admit orthogonal transformations $\mathcal{O}$ where $\mathrm{det}[\mathcal{O}]=-1$ (i.e.\ incorporating a reflection).
This type of dynamic is trickier to associate with a Hamiltonian, since it does not directly correspond to a generator.
Nonetheless, a recipe that allows us to do this is not mathematically inconceivable, although it is not so physical. 
For instance, once could assign some (connected) generator of $\SO{n}$ from the Hamiltonian, and then generate the finite-evolution $M$ by supplementing the exponential of the connected generator with a sign-flip on one element of the state vector.

To summarize: if we wish to have dynamics around a non-continuous symmetry in a state space, then we can restrict $t$ in Eq.~\eqref{eq:ExpMap} to a set of regular discrete values where $M(t)$ is an allowed\footnote{{\em Allowed} means that the transformation maps all states to states and satisfies any other constraints that are part of the theory.} transformation. 
This would change the nature of time in such theories to be a discrete quantity.

The implications of {\bf INV} are not affected by restricting dynamics to a discrete set of transformations.
Any discrete rotation can be produced from the same set of generators as a continuous rotation -- and one might still consider the same set of constraints on the generators and their relationship with energy observables.
If reflections are incorporated in a theory's dynamics,
 the set of reflections which obey {\bf INV} can be reduced to those which map $\effect{H}$ to $\effect{H}$ from an equivalence $\effect{H} \cdot (M\vec{\rho})=(\effect{H}M^{-1}) \cdot \vec{\rho}$.

Finally, desideratum {\bf QUAN} is not contradictory with having a formulation that allows for discrete time dynamics when presented with state spaces that have discrete symmetries.
The symmetry of quantum state spaces means that following a fixed prescription on a quantum state space should in general lead to continuous time behaviour; but that same prescription might yield discrete behaviour when presented with (non-quantum!) state-spaces.
Nonetheless, there are contexts where even within the laws of quantum mechanics such discrete time behaviour might be engineered. 
For instance, consider standard quantum dynamics, but explicitly only allowing the state to be written down at regular intervals chosen such that the number of pure quantum states (given a particular initial state) is a finite set. 
In such a context, the Hamiltonian would still have all its usual quantum meanings.

\section{Defining Hamiltonians for 3-dimensional state-spaces}
\subsection{Recipe for determining Hamiltonian}
\label{sec:RecipeH}
We now explicitly consider the case where the normalised states can be expressed by $3$-dimensional real vectors (and time-evolution preserves the normalization of these states)---such as the qubit Bloch sphere, or various popular alternative toy theories.
In this case there is a natural and powerful recipe for deriving a Hamiltonian from observations.

We shall make reference to a standard set of infinitesimal generators.  
The infinitesimal generator $G$ of the orthogonal group must satisfy the anti-symmetric constraint $G_{ij} = -G_{ji}$.
It can be expressed as some 
 linear combination of
\begin{equation}
 L_{\bold{x}} = \begin{bmatrix}0&0&0\\0&0&-1\\0&1&0\end{bmatrix} , \quad
 L_{\bold{y}} = \begin{bmatrix}0&0&1\\0&0&0\\-1&0&0\end{bmatrix} , \quad
 L_{\bold{z}} = \begin{bmatrix}0&-1&0\\1&0&0\\0&0&0\end{bmatrix}.
\end{equation}

In other words, any infinitesimal generator $G$ of $\SO{3}$
 can be written as a linear combination $\alpha_xL_x+\alpha_yL_y+\alpha_zL_z$, for $\alpha_x, \alpha_y, \alpha_z \in \mathbb{R}$.
This follows from the fact that the set of infinitesimal generators for the orthogonal group is precisely that set of matrices satisfying $G_{ij}=-G_{ji}$ and the above matrices are a basis for that set. 

{\bf The Recipe:}
{\em  The Hamiltonian for a given transformation is simply that vector of coefficients, i.e.\ $H_1=\alpha_x$, $H_2=\alpha_y$ and $H_3=\alpha_z$. 
That is,
\begin{equation}
\label{eq:Hdef3D}
G=H_1L_x+H_2L_y+H_3L_z,
\end{equation}
where $G$ is the generator, the $H_i$ are the components of the Hamiltonian and the $L$'s are the matrices above. 
This defines $\effect{H}$ for a given $G$.}

This was designed to satisfy the desiderata. 
In particular the choice of  $L_{\bold{x}}$,  $L_{\bold{y}}$ and $ L_{\bold{z}}$ as the basis guarantees that {\bf INV} is satisfied as shown below. 
The following table shows this is indeed the case. 
\begin{center}
\begin{tabular}{  r  >{\centering}m{0.8cm}  m{5cm} }
	\hline
    	Desid. & OK? & Why \tabularnewline
	\hline
   	{\bf OBS} & $\checkmark$ &  $\effect{H}$ is a real vector of the correct dimension.
 \tabularnewline
\hline
    	{\bf GEN} & $\checkmark$ & $G=H_1L_x+H_2L_y+H_3L_z$ where $G$ belongs to the Lie algebra of $\SO{3}$. \tabularnewline
\hline
    	{\bf INV} & $\checkmark$ & {\em See below.}  \tabularnewline
\hline
	{\bf QUAN} & $\checkmark$ & 
As we are using the structure constants of the orthogonal group for the generators in question (corresponding here to those of $\SU{2}$), substituting Eq.~\eqref{eq:Hdef3D} for $g_{ijk} H_k$ in Eq.~\eqref{eq:generalev} implies the quantum dynamics of Eq.~\eqref{eq:crossprod}. 
 \tabularnewline
	\hline
\end{tabular}
\end{center}

To demonstrate that {\bf INV} is satisfied:
From Eq.~\ref{eq:generalev},  $g_{ijk}$ for three-dimensional systems is determined by a single non-zero value $g_{123}$ and the associated anti-symmetrisations, which can be absorbed into a rescaling of $\effect{H}$
and allows us to specifically consider the structure constants of $\SO{3}$ associated with the choice of basis $\{ L_x , L_y, L_z \}$, which is given by the Levi-Civita symbol.
The equations of motion for a general state $\rho = \left(\rho_1, \rho_2, \rho_3\right)^\mathrm{T}$ are then
\begin{align}
	\dot{\rho}_1 &= \rho_2 H_3 - \rho_3 H_2\\
	\dot{\rho}_2 &= -\rho_1 H_3 + \rho_3 H_1\\
	\dot{\rho}_3 &= -\rho_2 H_1 + \rho_1 H_2,
\end{align}
and it follows that $\frac{d}{dt}\left(\effect{H}\cdot\rho\right)=\effect{H}\cdot\dot{\rho} = 0$.
 in general the exact resolution of $\effect{H}$ may not be possible from a single initial state.

If unnormalized states are considered, we must explicitly include the $\rho_0$ component of the state and the associated $H_0$ in the Hamiltonian.
The effect on $\effect{H} \cdot \vec{\rho}$ for any normalised state is to add a constant $H_0$ to the energy, which can be associated with the ``zero'' of energy relative to some absolute scale.
(In quantum mechanics this is seen in the Hamiltonians $H+\kappa\ident$ ($\kappa \in \mathbb{R}$) which give the same dynamics for all $\kappa$).
This $H_0$ has no direct dynamical consequence in theories where time-evolution preserves states' normalization. 
If the dynamics were affected by $H_0$, then $\rho_0$ would also have to change through the anti-symmetry of $g_{ijk}$ demanded by {\bf INV}.
Hence $H_0$ may be chosen arbitrarily in this recipe.

\begin{figure*}[thb]
\begin{centering}
\begin{subfigure}[t]{0.275\textwidth}
\includegraphics[width=\textwidth]{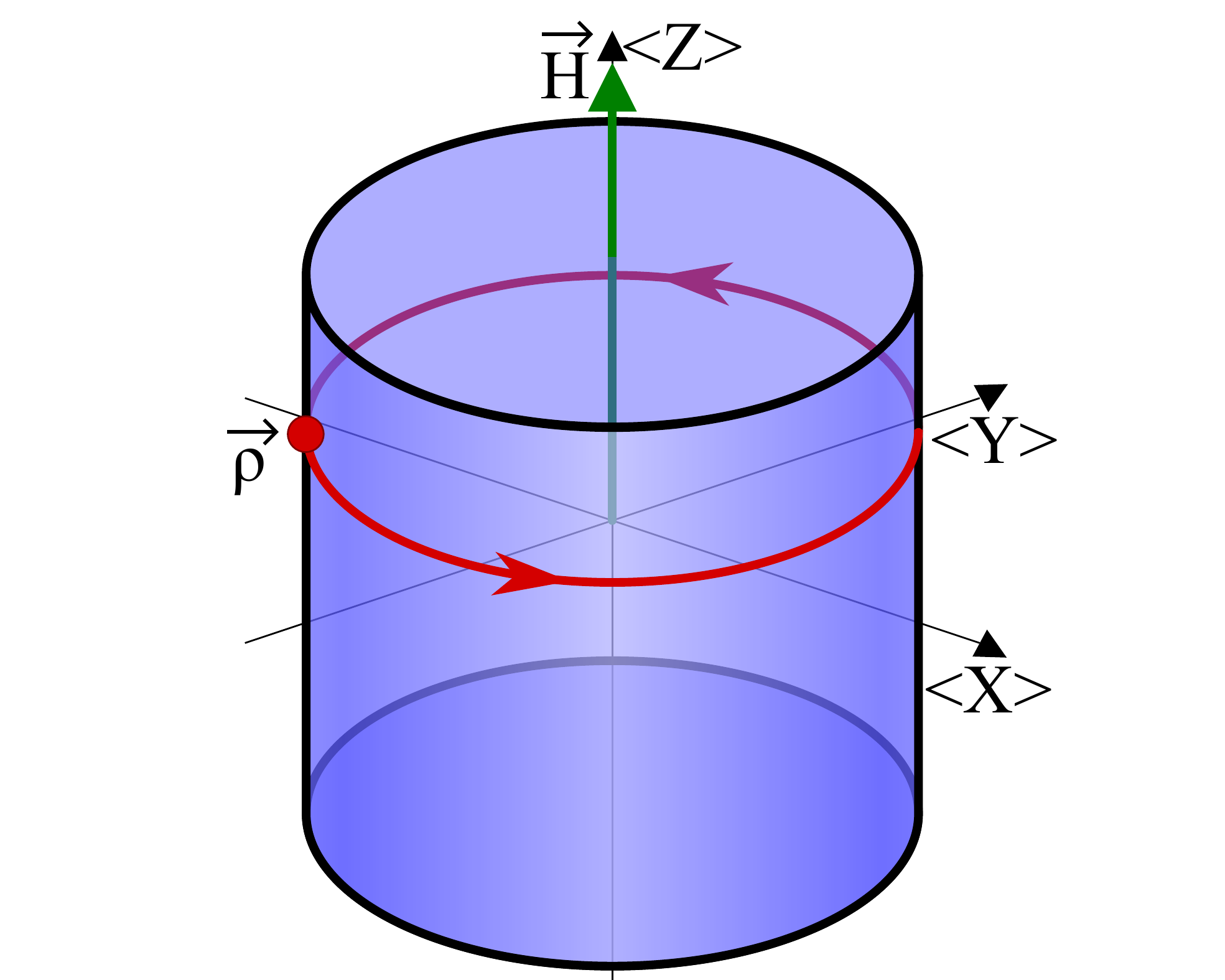}
\subcaption{Cylindrical state space. Arbitrary rotations around $Z$ axis permissible (shown); or $2$-fold rotations about any axis in $XY$ plane.
\label{fig:eg:cyl}
}
\end{subfigure}
\hspace{3em}
\begin{subfigure}[t]{0.275\textwidth}
\includegraphics[width=\textwidth]{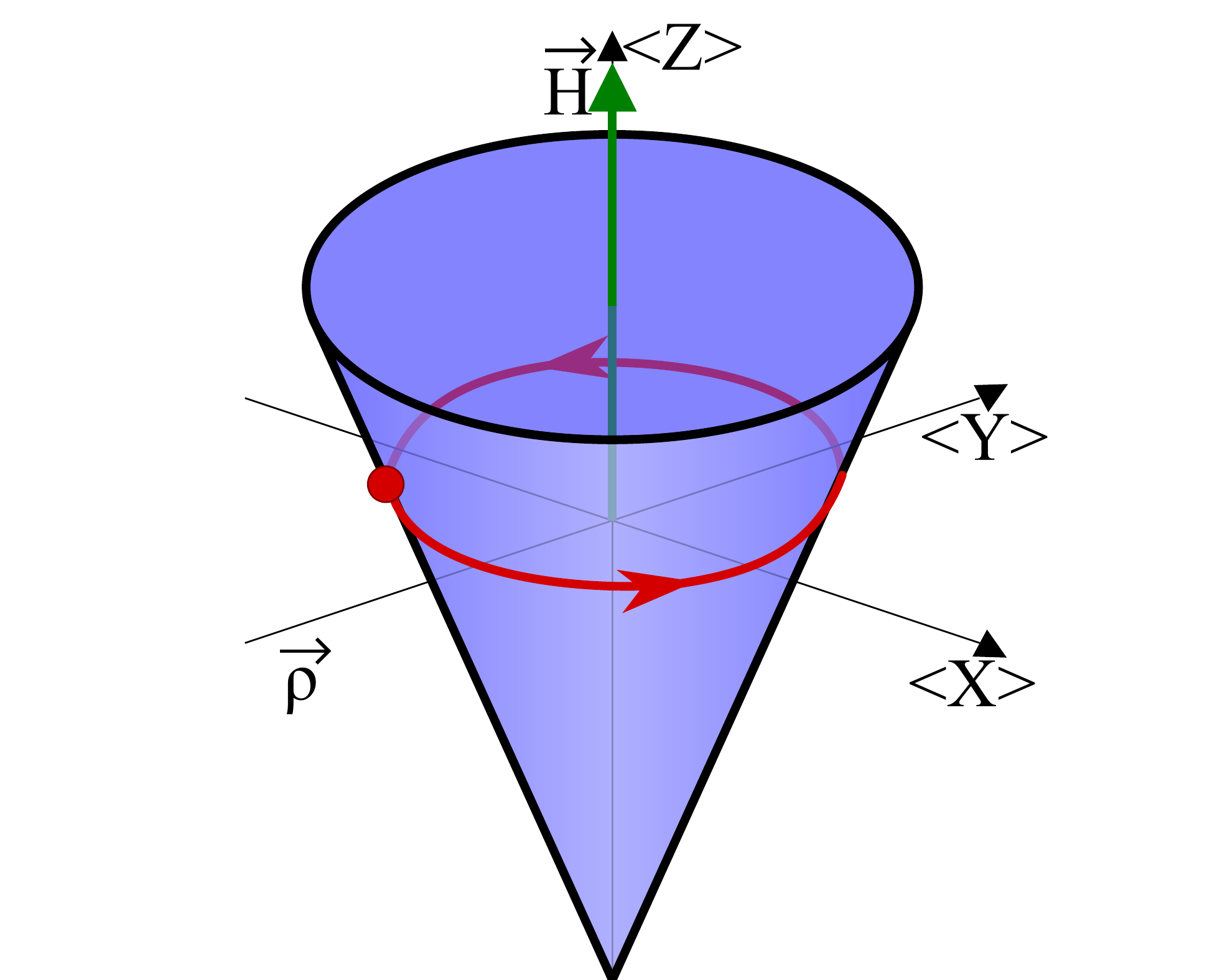}
\subcaption{Conic state space. Arbitrary rotations around $Z$ axis permissible.
\label{fig:eg:cone}
}
\end{subfigure}
\hspace{3em}
\begin{subfigure}[t]{0.275\textwidth}
\includegraphics[width=\textwidth]{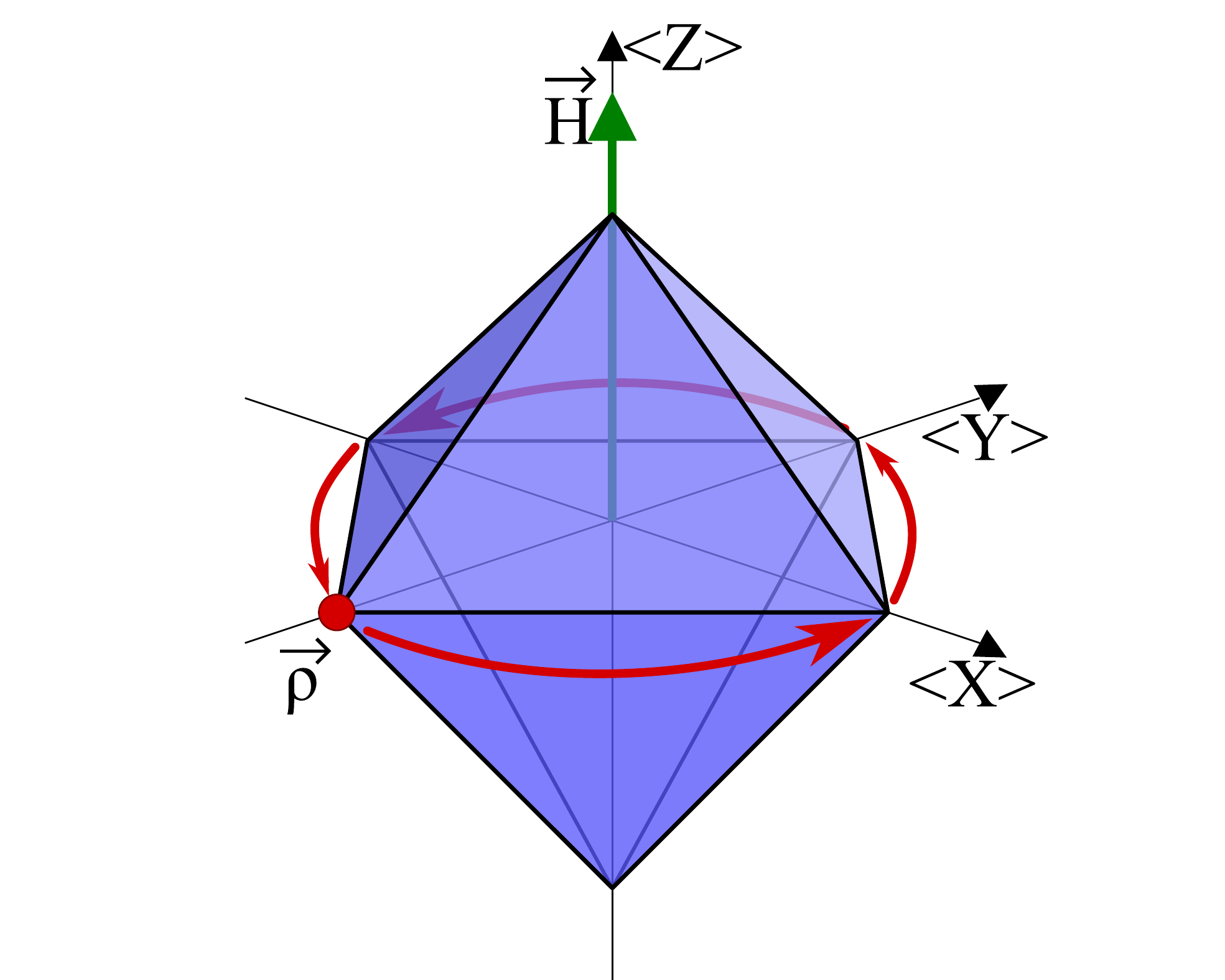}
\subcaption{Stabilizer states, Hamiltonian associated with dynamics around $Z$ axis.
\label{fig:eg:stab1}
}
\end{subfigure}
\\
\begin{subfigure}[t]{0.275\textwidth}
\includegraphics[width=\textwidth]{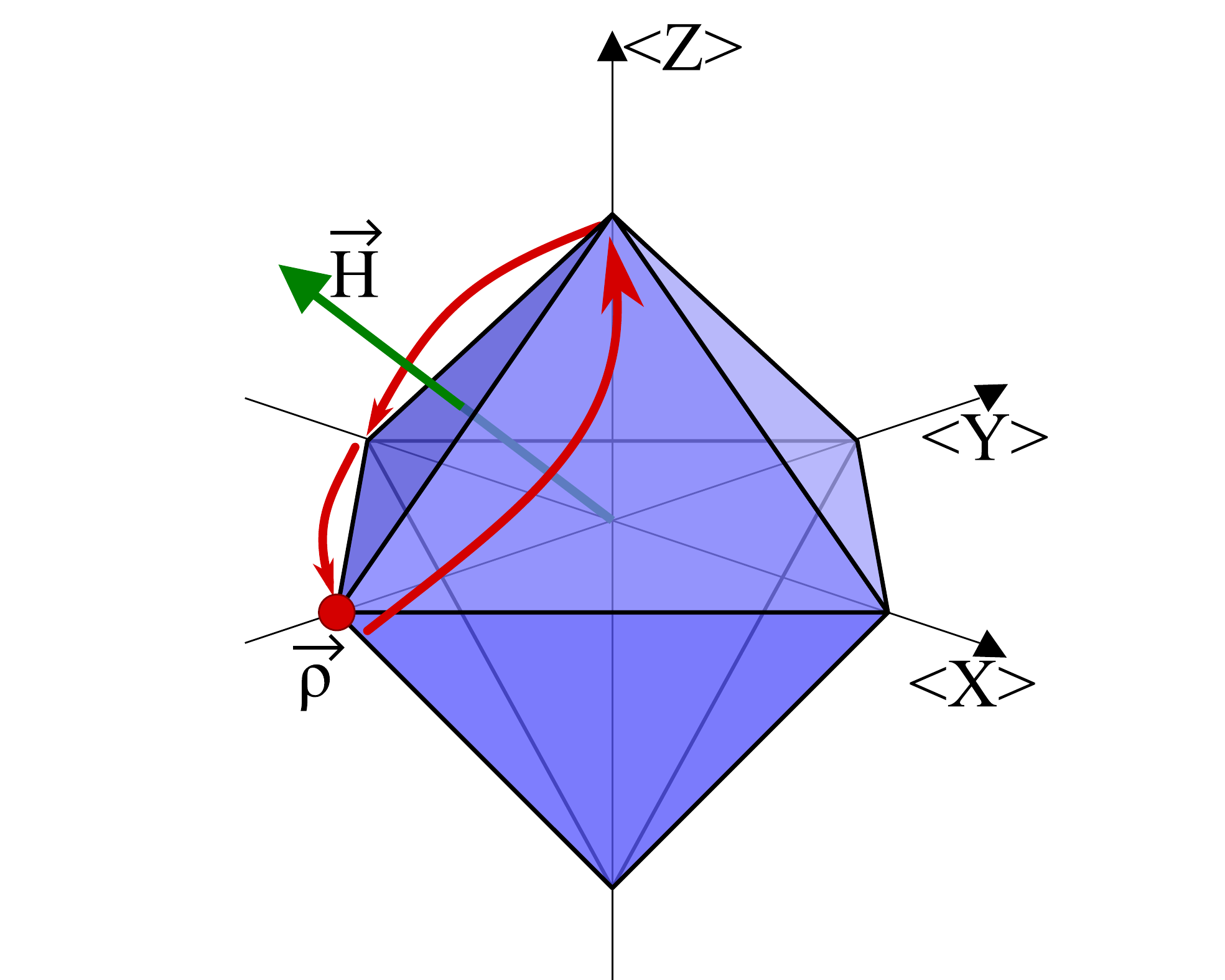}
\subcaption{Stabilizer states, Hamiltonian associated with dynamics with 3-fold symmetry.
\label{fig:eg:stab2}
}
\end{subfigure}
\hspace{3em}
\begin{subfigure}[t]{0.275\textwidth}
\includegraphics[width=\textwidth]{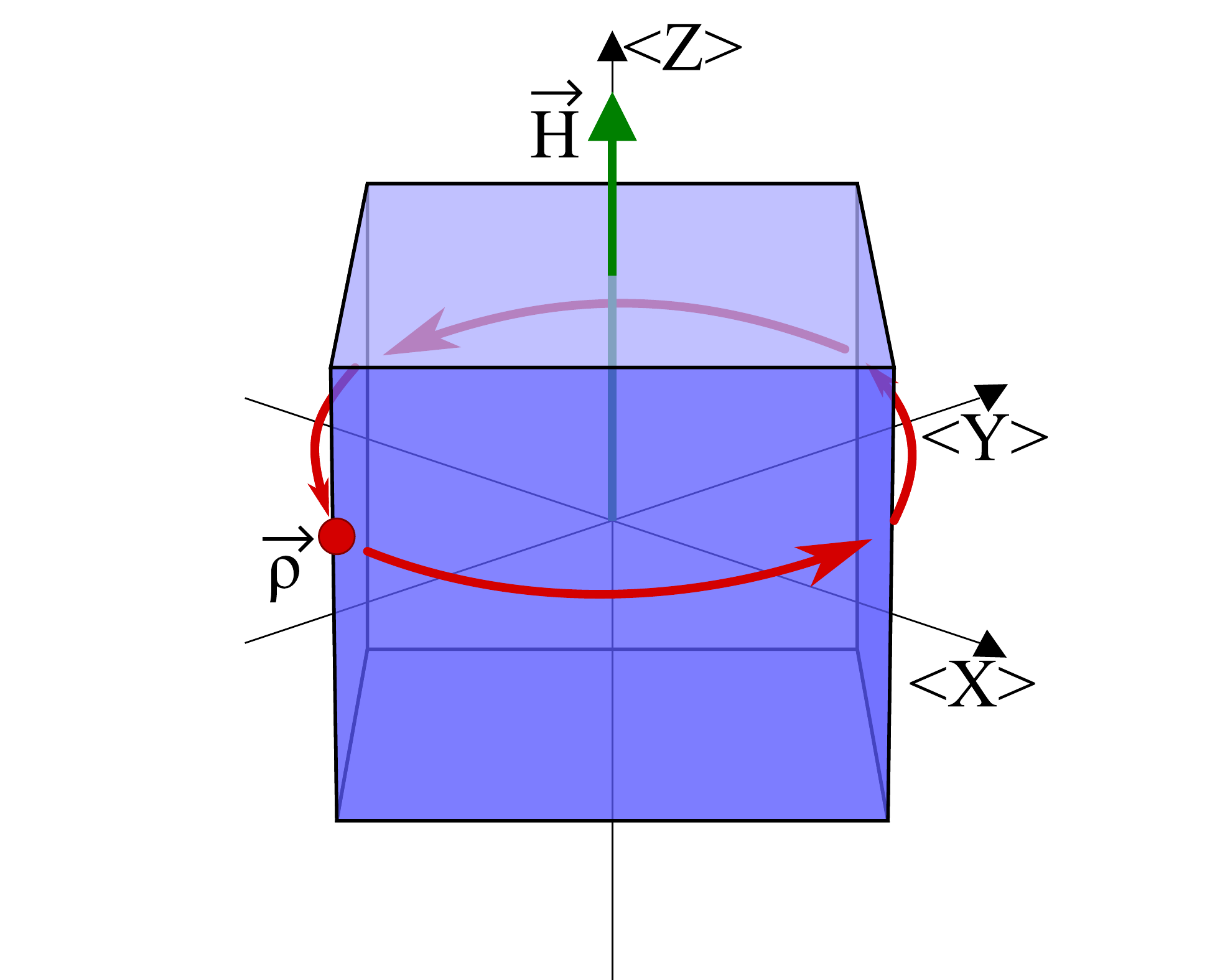}
\subcaption{Gbit with $3$ binary measurements.
\label{fig:eg:gbit}
}
\end{subfigure}
\hspace{3em}
\begin{subfigure}[t]{0.275\textwidth}
\includegraphics[width=\textwidth]{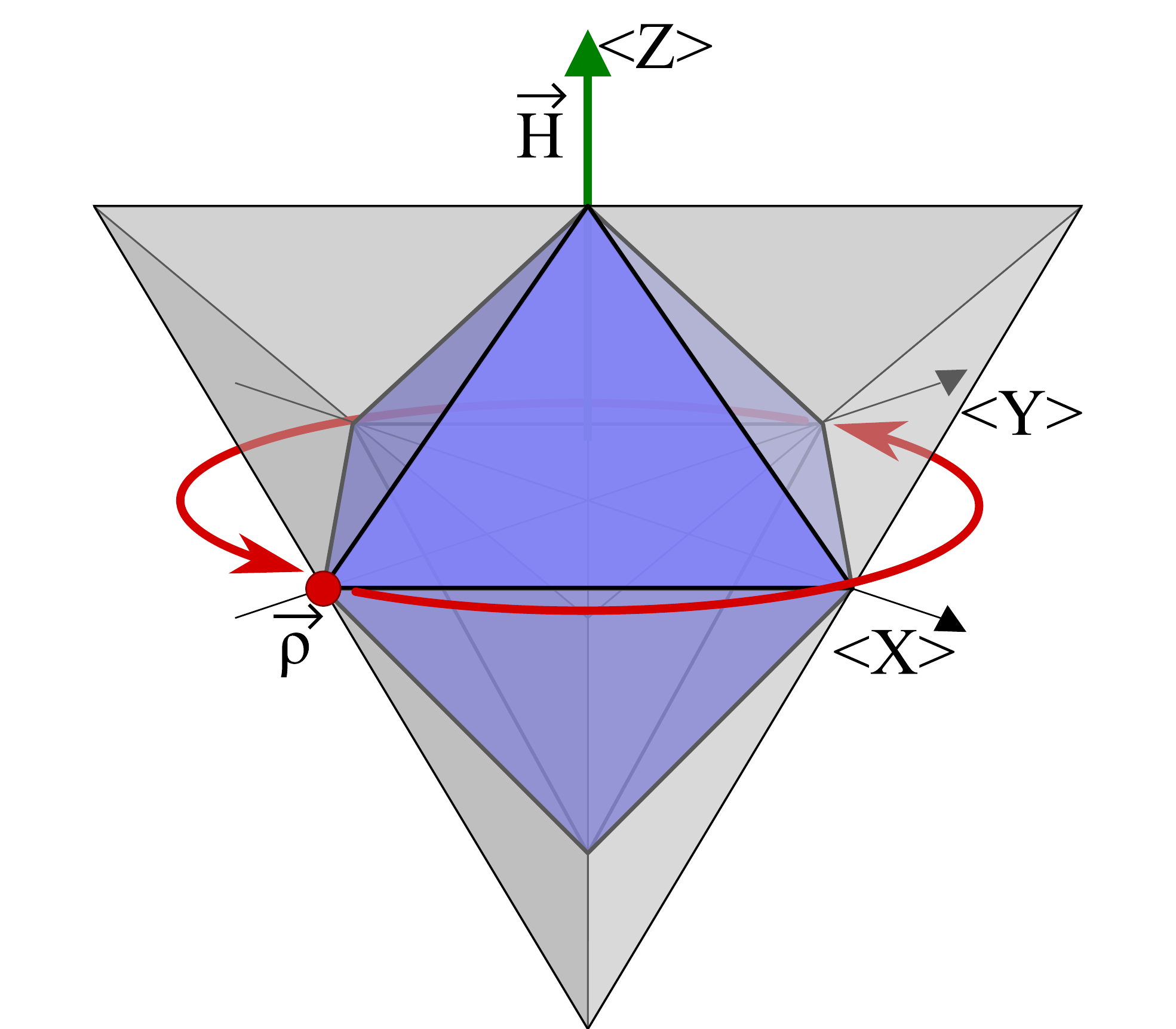}
\subcaption{Spekkens' toy model. The state space is octahedral, but the transformations must obey tetrahedral symmetries.
\label{fig:eg:spek}
}
\end{subfigure}
\caption{{\bf State spaces and Hamiltonians in example $3$-dimensional theories.} We express $\vec{\rho} = \left(\expect{X}, \expect{Y}, \expect{Z}\right)^{\rm T}$.
In all cases, the Hamiltonian acts as the axis of rotation.}
\label{fig:examples}
\end{centering}
\end{figure*}

\subsection{Examples} 
\inlineheading{Example 1: Qubit}%
The evolution of states in the Bloch sphere, and how this relates to the Hamiltonian, is the same whether using this recipe or Eq.~\ref{eq:crossprod} as in Section~\ref{sec:quantum_real}. Indeed, this is true by definition of {\bf QUAN}.
This is drawn in figure~\ref{fig:rotate}.

\inlineheading{Example 2: Cylindrical states space}%
Consider a cylindrical state-space where the allowed states of $X$ and $Y$ measurements are subject to an uncertainty relation $\expect{X}^2 + \expect{Y}^2 \leq 1$, but the $Z$ measurement is unconstrained (drawn in figure~\ref{fig:eg:cyl}).
In the corresponding theory, the Hamiltonian supplied by our recipe could correspond to arbitrary rotations about the $Z$ axis.
On the other hand, one could also define a Hamiltonian in the $XY$ plane. 
In this case, however, infinitesimal rotations are not valid transformations: only $2$-fold rotations (i.e.\ flipping the cylinder over) are allowed.

\inlineheading{Example 3: Conic states space}%
Consider a theory where states are restricted $\expect{X}^2 + \expect{Y}^2 \leq \frac{1}{2}\left(1+\expect{Z}\right)$. 
Such a constraint leads to an unusual conical state space (drawn in figure~\ref{fig:eg:cone}), where there is only one state where $\expect{Z}=-1$, but a family of states (with values $\expect{X}^2+\expect{Y}^2\leq1$ when $\expect{Z}=1$.
Like with the cylinder, a Hamiltonian in the $Z$ direction will accomodate arbitrary rotations.
However, since the symmetry in $Z$ is broken, the other discrete rotations are no longer allowed, and so this is the only Hamiltonian in the theory that can lead to dynamics.

\inlineheading{Example 4: Stabilizers}%
The convex hull of the stabilizer states (eigenvectors of Pauli matrices) form an octahedron sitting inside the Bloch sphere (touching where the axes cross through the sphere)~\cite{gottesman_class_1996}.
This can also be rotated, but only along certain directions and at certain angles. 
Each such evolution can be associated with a Hamiltonian with the above recipe.
(E.g.,\ see the 4-fold rotation in figure~\ref{fig:eg:stab1}, or the 3-fold rotation in figure~\ref{fig:eg:stab2}).

\inlineheading{Example 5: Box-world}%
As our first non-quantum example, let us consider the theory in which there are are three complementary binary measurements (without effects for joint measurements) but no uncertainty relation: that is, the {\em $3$-in-$2$-out gbit}~\cite{barrett_information_2007} of {\em box-world}.
This theory corresponds to a cubic state space.

Here, once again, the Hamiltonian is the axis of rotation.
However, unlike quantum theory, only certain axes and certain angles are allowed, as dictatated by the cubic symmetries,
 that is, the $4$-fold symmetries around the three ``natural'' observables (e.g.\ as drawn in figure~\ref{fig:eg:gbit}), but also the four $3$-fold symmetries through opposite vertices of the cube.
This would suggest that a universe whose state spaces are given by gbits can only admit discrete time evolution.

\inlineheading{Example 6: Spekkens' toy model}%
Spekkens' toy model~\cite{spekkens_evidence_2007} is a popular toy theory that exhibits many features of quantum theory.
Here, one takes a classical variable specified by $n$ bits of information, and places the epistemic restriction that one can only know $n/2$ bits at any time.
The simplest single system in this framework corresponds to a $2$-bit system, where one knows the value of either of the bits, or alternatively the correlation between them. 
This also corresponds to a system with three measurements each with two outcomes.

Although {\em a priori}, Spekkens' toy model is not a GPT, it can be closed into a convex state space and represented as a theory within the convex framework~\cite{vanenk_toy_2007,janotta_2013,garner_framework_2013}.
The state space then is an octahedron, much like the stabilizers~\cite{pusey_stabilizer_2012}.
However, the allowed transformations are different, as they must correspond to valid operations on the underlying $2$ bits---that is, a four-level classical system, with tetrahedral symmetry. 
(For instance, it is impossible to flip the value of just one bit without also flipping the correlation; and this constrains the symmetries we might see).

Thus, when we apply our Hamiltonian recipe to this system, we find once more that we are limited to discrete times (figure~\ref{fig:eg:spek}).
The 3-fold symmetries of the stabilizers are still present; but the rotations about the principle axes are limited to $0$ or $\pi$.

\subsection{Higher dimensional case}
We now consider how to define Hamiltonians in higher dimensions. 
Naturally this becomes more difficult to visualise, yet may allow for an even richer set of non-quantum theories.
Indeed, for systems with a three-dimensional state space,
 the anti-symmetry of $g_{ijk}$ in Eq.~\eqref{eq:generalev} permits only permutations of a single triple to be non-zero (and it must be non-zero to avoid trivial dynamics), 
 forcing a greater equivalence between quantum and non-quantum theories than may be present in the higher dimensional case.

The generators $L_X$, $L_Y$ and $L_Z$ naturally generalise to higher dimensions by noting that they are the basis for skew-symmetric matrices.
However, if one uses the structure constants for the orthogonal group naively as a method for determining $\effect{H}$ in the same way as we did for the 3D case, there is the problem of dimensional mismatch.
There is on the order of $d^2$ such generators whereas the vector for an observable should be $d$-dimensional.
In the quantum case this is not a problem because one uses a restricted generator set---not the full orthogonal group set---with the associated $\SU{n}$ structure constants.
In order to have a well-defined Hamiltonian as both observable and generator in general, one possible route is accordingly to accept a reduced generator set.
Otherwise one might consider multiple Hamiltonians for a given generator. 

\subsection{Classical mechanics}
While it is beyond the scope of this paper to undertake an in-depth analysis of the classical mechanical Hamiltonian from a GPT perspective, we note that the classical mechanical time evolution fits into the paradigm used here.
To see classical mechanical evolution as an orthogonal transformation on a real vector use the Liouville equation:
\begin{equation}
- \frac{\rm d}{{\rm d}t}\rho =\{\rho, H\}:=\mathcal{L (\rho)},
\end{equation}
where
\begin{equation}
\mathcal{L (\cdot)}=\sum_i \frac{\partial H}{\partial p_i}\frac{\partial (\cdot)}{\partial q_i} - \frac{\partial H}{\partial q_i}\frac{\partial (\cdot)}{\partial p_i}
\end{equation}
is the Liouville operator.
The real vector here is the phase space density $\rho(x_1,p_1,x_2,p_2,...)$.
The phase space density can by assumption be expanded in some basis of orthogonal functions $\{\phi_i\!\left(x,p\right)\}_i$ (e.g.\ Fourier series plane waves or delta functions).
One can prove through standard arguments that the operator $\mathcal{L(\cdot)}$ is antisymmetric with respect to swapping basis element vectors, assuming that the phase space of interest is bounded so that the basis functions vanish at the boundaries, and that H is of the form $\frac{p^2}{2m}+V(x)$ (see e.g.\ \cite{tuckerman_lecture_2002}):
\begin{equation}
\mathcal{L}_{kl}=-\mathcal{L}_{lk},
\end{equation}
where $\mathcal{L}_{kl}=\int {\rm d}x\,{\rm d}p \,\phi_k (x,p) \mathcal{L}(\phi_l (x,p))$.
Thus the time evolution, for a constant generator is
\begin{equation}
\rho (t)= e^{\mathcal{L}t}\rho (0)
\end{equation}
and $e^{\mathcal{L}t}$ is an orthogonal matrix (as the matrix exponential of an antisymmetric matrix is an orthogonal matrix).
We also note that the quasi-probability Wigner function can act as the real vector in question, fitting into the GPT framework used here~\cite{lin_tunnelling_2016}.

\section{Energy eigenstates and phase transformations}
\label{sec:Phase}
One important feature of Hamiltonians in quantum theory is that the {\em eigenstates} of a system's Hamiltonian (referred to as {\em energy eigenstates})
provide a special basis for writing down the dynamics of the system in a particularly simple manner.
Namely, in the standard picture of quantum theory, time evolution induces variations of phases in front of energy eigenstates. 
This is, for some pure quantum state $\ket{\psi} = \sum_i \alpha_i \ket{e_i}$ (where $\ket{e_i}$ are the energy eigenstates and $\alpha_i$ are complex coefficients such that $\sum_i |\alpha_i|^2 =1$), Hamiltonian evolution for time $t$ changes the state to 
\begin{equation}
\label{eq:QuantumPhaseEnergy}
\ket{\psi'} = \sum_i \exp\left(-\frac{i E_i t}{\hbar}\right) \ket{e_i},
\end{equation} where each $E_i$ is the eigenvalue associated with eigenstate $\ket{e_i}$.

Beyond defining the Hamiltonian in GPTs, 
 it is interesting to consider the extent to which the evolution determined by our GPT Hamiltonians resembles that of a phase transformation (in the manner of \citet{garner_framework_2013}), 
 as this has the potential to inform understanding of decoherence and thermalisation-type effects.
To generalise this beyond quantum theory, we will need a general definition of phase.

\subsection{Phase in the convex framework}
Consider a pure quantum state $\ket{\psi} = \sum_i c_i e^{i \phi_i} \ket{i}$ expressed in the basis of $\{\ket{x_i}\}$.
Here, the measurement statistics associated with $\{\ket{x_i}\}$ are given by the probabilities $p_i = |\braket{x_i | \psi}|^2 = |c_i|^2$,
 and one can see that the values $\{\phi_i\}$ can be freely changed without changing these statistics.
As such, $\{\phi_i\}$ are here referred to as {\em phases} associated with the measurement.
One can then take an active picture of phase, whereby one considers the set of transformations on a state that preserve the statistics of a given measurement.
For reversible quantum operations, this is the subgroup $G_\Phi := U(1)\oplus\ldots U(1) \subset U(n)$.
By assigning a ``reference state'' for a given set of statistics, (e.g.\ a natural choice is $\ket{\psi_0} = \sum_i c_i \ket{x_i}$), one can see there is a one-to-one relationship between elements of this subgroup and choices of phases $\{\phi_i\}_i$, namely $\ket{\psi} = g \ket{\psi_0}$ for $g\in G_\Phi$.

This particular group-theoretic notion of phase was generalised to the convex framework in~\citet{garner_framework_2013},
 where without the comfortable underlying structure of Hilbert spaces, the phase has been defined solely in terms of transformations.
A transformation $T$ is said to be a phase transformation with respect to a measurement $B$ made up of effects $\{\effect{e_b}\}$, if $\effect{e_b}\cdot\vec{\rho} = \effect{e_b}\cdot T \vec{\rho}$ for every $b$, and every state $\rho$ in the theory.
That is, a phase transformation with respect to a measurement always preserves the statistics of that measurement. 
If one considers this constraint on the group of reversible transformations, we find that it induces a particular sub-group deemed the {\em phase group} associated with measurement $B$.

\subsection{States with well-defined energy}
\label{sec:energyeigenstates}

\begin{figure}[tbh]
\centering
\includegraphics[width=0.35\textwidth]{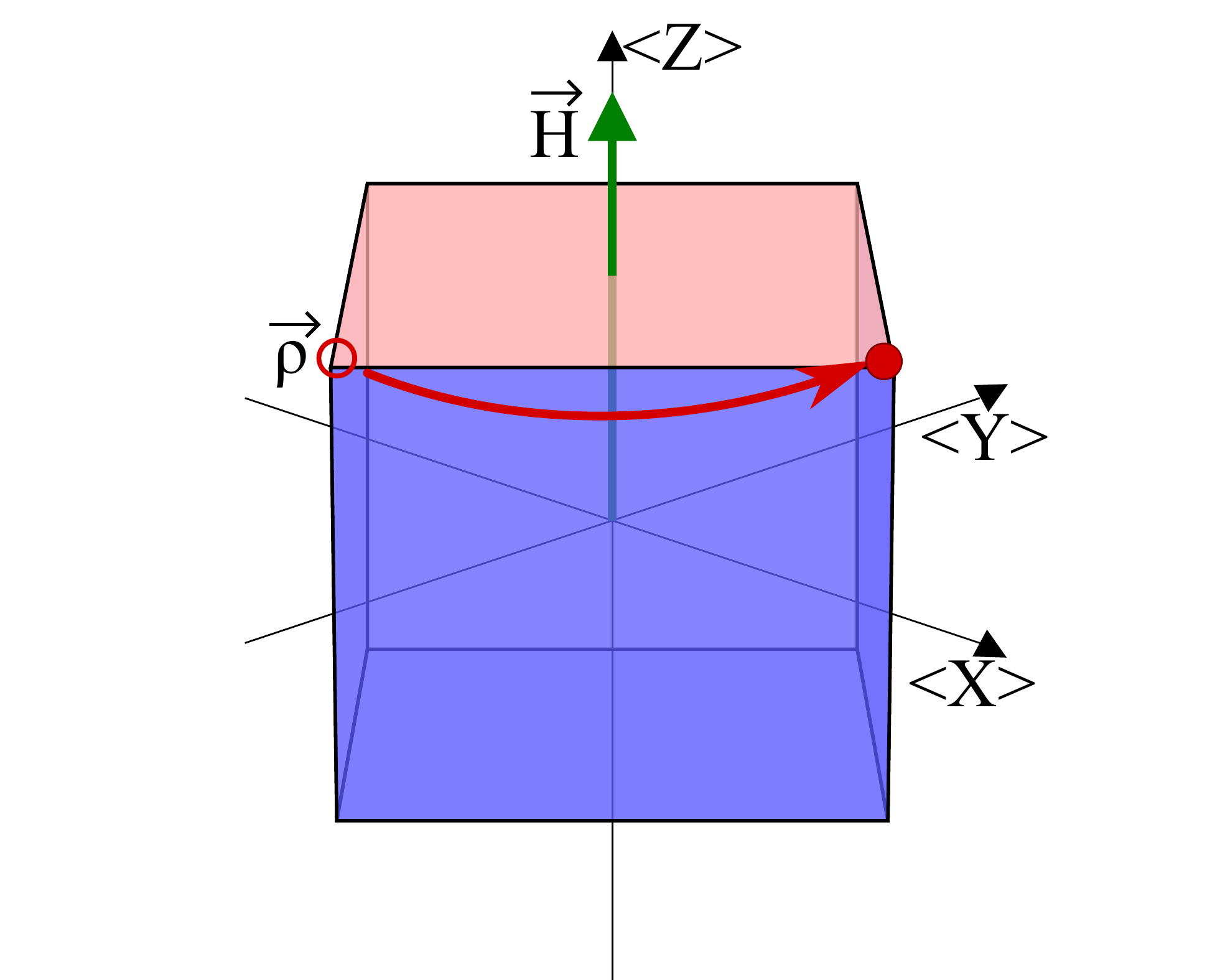}
\caption{
{\bf Well-defined energy does not imply stationarity under time-evolution.}
The entire top plane (shaded red) of the cube is a state with a definite energy. Nonetheless, not every state on this plane is invariant under time evolution, as shown by the red example state $\rho$.
}
\label{fig:BoxWorldEigen}
\end{figure}

The notion of time evolution putting a phase in front of energy eigenstates is more subtle in general theories than it is in quantum theory.
Consider an operational definition of a state having {\em well-defined energy}: 
 let there be some energy measurement with effects $\{\effect{h_i}\}$.
A state $\vec{\rho}$ has well-defined energy (namely is said to be in energy level $i$) if $\effect{h_i}\cdot\vec{\rho} = 1$ (and by implication $\effect{h_j}\cdot\vec{\rho} = 0$ for all other $j\neq i$).

However, unlike in quantum theory, the states with well-defined energy are not necessarily stationary under time evolution.
Consider a cubic state space such as that of the gbit with Hamiltonian and evolution as in Fig.~\ref{fig:BoxWorldEigen}.
The top and bottom planes of the cube have well-defined energy (and this is not changed by the evolution) but the states themselves are not stationary under time evolution. 
Instead, we can talk of convex sets of states within planes which are stationary with respect to the Hamiltonian observable under time evolution, such as the lower and upper plane of the cube. 

On the other hand, there are some special cases of theories where being in a definite energy state does guarantee stationarity under time evolution due to the uniqueness of the state. 
This corresponds to theories where the energy observable is a spectral measurement~\cite{krumm_thermodynamics_2017}.
Noteable examples of this include complex and quaternionic quantum theory~\cite{graydon_2011}.
As such, a postulate that states of definite energy do not change under time evolution could be taken as an axiom towards selecting quantum theory, or indeed as an alternative physical motivation that implies existing axioms such as postulates $1$ and $2$ of~\citet{barnum_higher-order_2014}.

\subsection{Hamiltonian dynamics as phase transformations}
Is the time evolution a phase transformation with respect to the Hamiltonian measurement? 
In quantum theory, Eq.~\eqref{eq:QuantumPhaseEnergy} shows clearly that it is.
In general, however, a cursory glance tells us that {\bf INV} places a single constraint on transformations, whereas restricting oneselves to the set of phase-transformations of an $n$-outcome measurement effectively places $n$ constraints
As such, we may wish to state a stronger desideratum {\bf INV*} that mandates that {\em all} energy-related statistics of the decomposition of $\effect{H}$ should be invariant under Hamiltonian dynamics.
Under {\bf INV*}, it follows that Hamiltonian dynamics are phase transformations of an energy measurement.

While it is clear that {\bf INV} $\neq$ {\bf INV*} in general\footnote{
As a pathological example, consider a three level system where $E_1 = 0$, and $E_3 = 2 E_2$; a transformation from a state $\vec{\rho}$ with well-defined energy in $E_2$ ($\effect{e_2}\cdot\vec{\rho} = 1$) to $\rho'$ where $\effect{e_1}\cdot\vec{\rho'}  = \effect{e_3}\cdot\vec{\rho'} = 0.5$ would satisfy {\bf INV} but not {\bf INV*}.
}.
there are some conditions where {\bf INV} does imply {\bf INV*}.
Firstly, although restricting a transformation to be in a phase-group seemingly places $n$ constraints, if the transformations are also restricted to those that map normalized states to normalized states, we see that only $n-1$ of the phase group constraints are independent (because the probabilities of outcomes of each measurement must add to $1$).
As such, for systems with two different energy levels,
 we straight away see that {\bf INV} imposes that time-evolution is in the phase group of the Hamiltonian measurement.
This holds true for any theory where the energy measurement distinguishes at most between two outcomes, including any to which the recipe in Section \ref{sec:RecipeH} applies, $d$-dimensional balls ($2$-level Jordan algebras) and $d$-in $2$-out gbits.

This distinction between operations that conserving the expectation value of energy, and those that are diagonal with respect to the Hamiltonian (i.e.\ are phase operations of the enegy measurement) has also appeared in the very different context of quantum thermodynamics~\cite{goold_thermoreview_2016} -- where one often wishes to identify a set of energy-conserving ``thermal operations''~(see e.g.\ \cite{goold_thermoreview_2016}). 
The set of allowed thermal operations is different if {\em average} energy is preserved~\cite{skrzypczyk_thermo_2014} (analagous to {\bf INV}), vs.\ the more restrictive conditions that the thermal operation commute with the Hamiltonian~\cite{horodecki_2013} (analagous to {\bf INV*}).

\subsection{Determining the energy of a system}
For a theory with a given energy measurement $\{\effect{e_i}\}$, and a Hamiltonian $\effect{H}=\sum_i E_i\effect{e_i}$,
 the real numbers $\{E_i\}$ may be thought of as ``energy eigenvalues'': namely, they assign a definite energy value $E_i$ to any state $\rho_i$ with well-defined energy in level $i$.

In quantum theory, the choice of such numbers is not meaningless.
Consider a two-level system $\ket{\psi(0)} = \alpha_1 \ket{e_1} + \alpha_2 \ket{e_2}$, evolving under the Hamiltonian $H = E_1 \ketbra{e_1}{e_1} + E_2\ketbra{e_2}{e_2}$.
After time $t$ the system is in state $\ket{\psi(t)} = \alpha_1 \exp\left({-\frac{i E_1 t}{\hbar}}\right) \ket{e_1} +  \alpha_2 \exp\left({-\frac{i E_2 t}{\hbar}}\right) \ket{e_2}$.
Taking into account global phases,
 we see that there is some $\tau$ such that $\ket{\psi(\tau)} = \ket{\psi(0)}$, namely when 
$\tau = {h}/{\left(E_2 - E_1\right)}$.
As such, the difference in energy values assigned to each eigenstate determines the speed of the evolution.

Consider theories with Hamiltonians given by the recipe in Section~\ref{sec:RecipeH}: namely those where the maximal measurement~{\cite{garner_phase_2015} distinguishes between two mutually exclusive possibilities and states can be described by $3$-dimensional real vectors (i.e.\ the qubit and its foils, as per our earlier examples).
In this case, the Hamiltonian observable should have the form
\begin{equation}
\label{eq:TwoLevelH}
\effect{H}=E_1\effect{e_1}+E_2\effect{e_2}.
\end{equation}

In such theories, since the dynamics are guaranteed to be a simple rotation, we can determine the period of motion and operationally find $E_2 - E_1 = \frac{1}{\tau}$, where $\tau$ is set by the experiment (note, we have set our units of energy such that $h = 1$).

If we then further impose that the physics should not care about a general shift in energy (that is, the dynamics should be a function only of relative energies), then this allows us to set $E_1=0$ and write
\begin{equation}
\label{eq:UniqueH}
\effect{H}=\frac{2\pi}{\tau}\effect{e_2}.
\end{equation}
The energy observable associated with any given dynamic can thus be determined for this class of theory.

What about cases where the energy measurement has three or more outcomes?
Here, an additional conceptual difficulty arises: namely, which aspects of the dynamics should be assigned to which energy levels.
[In general, theories do not have an underlying Hilbert space structure which leads to the obvious form of Eq.~\eqref{eq:QuantumPhaseEnergy}.]

A principle for associating a given phase transformation with a given state or set of states was proposed in~\cite{dahlsten_uncertainty_2014} in the setting of generalized interferometry experiments.
Here, the key degree of freedom (conventially labelled $Z$) describes a choice from a set of disjoint paths, or {\em branches} (e.g.\ the ``which path'' measurement on the two arms of a Mach--Zehnder interferometer).

In such theories, each branch $i$ is associated with an effect $\effect{e_i}$ (such that the total set of effects over all branches forms a measurement),
 and a state $\vec{\rho}$ is said to be localized to a branch $i$ if $\effect{e_i}\cdot\vec{\rho}=1$ for that branch, and $\effect{e_j}\cdot\vec{\rho}=0$ for all $j\neq i$.
One can then talk about the localization of transformations to branches using the principle of {\em branch locality}~\cite{dahlsten_uncertainty_2014}:
 if a state has no support on branch $i$, then any transformation that effects a change in a state $\vec{\rho_{\lnot i}}$ where $\effect{e_i}\cdot\vec{\rho_{\lnot i}} = 0$ cannot be said to be localized to branch~$i$.
This statement allows one to induce a hierarchical structure of subgroups of transformations that can and cannot be performed locally to (sets of) branches. 
For example (figure~\ref{fig:BL}), in a three-branch quantum system, branch locality would forbid operations taken on branch $1$ to induce a phase between branches $2$ and $3$~\cite{garner_2017}.

\begin{figure}[h]
\centering
\includegraphics[width=0.35\textwidth]{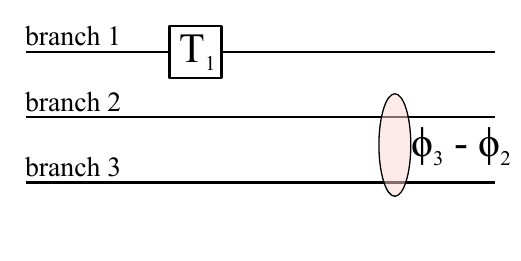}
\caption{
{\bf Branch locality in $3$-level quantum theory.}
Suppose there was an interferometry experiment with three spatially disjoint branches.
Any action taken on the top branch $T_1$ will not affect the relative phase (or indeed any other statistic) $\phi_3 - \phi_2$ between states on the bottom two branches (proof in appendix B of \citet{garner_2017}).
}
\label{fig:BL}
\end{figure}

Although this principle of branch locality was first motivated with respect to position measurement, 
 it is natural to consider the implications if the branch localization of transforms is also applicable to energy measurements. 

The principle of branch locality can be quite restrictive as to what transformations may be localized.
As an example, take the cubic gbit state-space (Fig.~\ref{fig:BoxWorldEigen}).
Here, the upper and lower planes represent the branches associated with definite outcomes of the $Z$ measurement.
For a given phase transformation to be localised to one of these planes it must leave the other invariant, and in the case of the gbit no such non-trivial transformation exists~\cite{dahlsten_uncertainty_2014}.
Thus we see that the time evolution in the cube example (rotation around the central axis) is not a composition of phase transformations localised to either energy branch, as that would be the identity operation.
Indeed, the unique assignment Eq.~\ref{eq:UniqueH} only worked because it is concerned with {\em relative phases}, and, in the two-level case, a single relative phase encapsulates {\em all phase dynamics}.
For higher dimensional systems in box-world, we run into problems, since there would now be phase dynamics between $3$ different pairings of branches (and possibly also tripartite phase dynamics),
 but branch locality tells us that all box-world dynamics must be global\footnote{
This does not mean that there are no phase dynamics -- take for instance the Aharonov--Bohm effect in quantum theory, whereby some global operation induces phases between branches. However, in quantum theory, a transformation with statitically identical action on the states can also be induced by putting pieces of glass on each individual branch; in box-world, a global transformation akin to the Aharonov--Bohm effect is permissible, whereas the analogous local construction that induces the same phase transformation is not possible.
}
 (strictly: a system with $d$ branches cannot be localised even to a subset of $d-1$ branches)~\cite{dahlsten_uncertainty_2014,garner_phase_2015}.

On the other hand, in the quantum case any phase transformation can be implemented as a local phase transformation~\cite{dahlsten_uncertainty_2014}, namely because there is only a single definite state with respect to each outcome of the phase measurement, and thus constraints on the state-space will prevent local operations from violating branch locality.
Likewise, quaternionic quantum theory also allows for this kind of branch localisation~\cite{garner_2017}.

With this in mind, we see that the dynamics such as that of the cylinder of Fig.~\ref{fig:eg:cyl} or the cube in Fig.~\ref{fig:eg:gbit},
 cannot be explained in a manner similar to Eq.~\eqref{eq:QuantumPhaseEnergy}, whereby the (generalized) evolution is driven by the independent application of phases between different energy ``eigenstates''.
Indeed, Hamiltonian dynamics consistent with principles of branch locality may be a signifier of quantum theory (or something structurally similar to  it), as we would expect from the results in \citet{barnum_higher-order_2014}.

Let us return to the question of determining energy values from a system's dynamics (if this is even possible).
Suppose we wish to determine relative energies between pairs of energy levels.
For theories where transformations can be branch localised to pairs of energy levels,
 we could consider the action of time evolution on the set of states supported on specific pairs of energy levels (say $i$ and $j$),
 and hope that every state in this set is mapped to itself after the same length of time $\tau_{ij}>0$.
If this is the case, then one can determine the relative energy $\Delta_{ij} := 1/\tau$.
From these $\Delta_{ij}$ one can infer a set of simultaneous equations $\Delta_{ij} = E_j - E_i$ to solve for the energies $\{E_i\}$ up to some constant offset.
The precise set of theories where these equations are consistent is beyond the scope of this current discussion, 
 but we suspect that if this does not outright select quantum theory, it will take us fairly close to it.

\subsection{Time and discrete evolution}
Finally, let us make a few remarks on the nature of time in theories with discrete evolution (for example, quantum stabilizers~\cite{gottesman_class_1996}, the gbit~\cite{barrett_information_2007}, or Spekkens' toy model~\cite{spekkens_evidence_2007}, as discussed above).
If we impose that we only measure time at discrete intervals (say, $\mathcal{\tau}$),
 in order for the generated dynamics to correspond to allowed rotations of the state-space, 
 one has to restrict the allowed relative energies to a discrete set such that rotation for time $\tau$ is always allowed.

Moreover, in such a theory not only are the allowed energies discrete, but the set of operationally distinct energies may be limited in number. 
Consider the $4$-fold rotation of the gbit around a principle axis. 
Unlike continuous time theories, where one can always consider evolution over a shorter time, in the discrete case rotation by $\pi/2$ and by $5\pi/2$ are operationally indistinguishable. 
Thus, if we look to the dynamics of this system to determine the energy eigenvalues, there would only be $4$ distinct energies. 
This relates to the phenomenon known as `aliasing'. 
Suppose we could only measure a rotating system at regular time intervals: there is in fact a discrete infinite family of frequencies that we might assign consistently with our readings.
In the case where time itself is discretized---rather than just taking a discrete subset of samples of a continuous reality---then these aliases become in fact an unphysical gauge freedom, since there would be no way to distinguish between them (e.g.\ by taking samples at a faster interval).

We thus draw the following conclusions for Section \ref{sec:Phase}:
 (i) time evolution is a phase transformation with respect to energy more generally too,
 (ii) in general, states with well-defined energy are not fixed under time evolution, unlike in quantum theory, and  
 (iii) the time evolution is not localisable to energy eigenstates in general. 
Points (ii) and (iii) may be important characteristics of quantum theory.

\section{Summary and Outlook}
We have seen how a definition of energy very similar to quantum theory can be extended in a concrete manner to a much more general set of theories. 
In particular, we focussed on 3-dimensional state spaces and found a definition of a Hamiltonian which satisfied four desiderata: 
 (i) {\bf OBS}: that the Hamiltonian is an observable, 
 (ii) {\bf GEN}: that it determines the generator of the time evolution,  
 (iii) {\bf INV}: That its expectation value is invariant under the time evolution, and 
 (iv) {\bf QUAN}: that the definition should be consistent with the quantum definition.
We investigated such Hamiltonians for certain example theories.
We moreover found that time evolution can also be seen as a phase transformation, but in general not without sacrificing the association of eigenstates as stationary states.

We anticipate that progress in defining and understanding Hamiltonians in this general framework will lead to progress in understanding energy related phenomena, including: (i) thermodynamics, (ii) the difference between quantum and classical time evolution, and (iii) tunnelling in post-quantum theories. 

\section{Acknowledgements}
We thank George Knee and Benjamin Yadin for useful comments.
We are grateful for financial support from the UK Engineering and Physical Sciences Research Council, the John Templeton Foundation, the Foundational Questions Institute, EU Collaborative Project TherMiQ (Grant Agreement 618074), the London Institute for Mathematical Sciences and Wolfson College, University of Oxford.

%

\end{document}